# Direct and diffuse shading factors modelling for the most representative agrivoltaic system layouts

Sebastian Zainali[1,*], Silvia Ma Lu[1], Bengt Stridh[1], Anders Avelin[1], Stefano Amaducci[2], Michele Colauzzi[2], Pietro Elia Campana[1,*]

[1]Mälardalen University. Dept. of Sustainable Energy Systems, Box 883, 721 23 Västerås (Sweden)

[2]Università Cattolica del Sacro Cuore, Dept. of Sustainable Crop Production, Emilia Parmense 84, Piacenza (Italy)

- Corresponding author: sebastian.zainali@mdu.se, pietro.campana@mdu.se

## Abstract

Agrivoltaic systems are becoming more popular as a critical technology for attaining several sustainable development goals such as affordable and clean energy, zero hunger, clean water and sanitation, and climate action. However, understanding the shading effects on crops is fundamental to choosing an optimal agrivoltaic system as a wrong choice could lead to severe crop reductions. In this study, fixed vertical, one-axis tracking, and two-axis tracking photovoltaic arrays for agrivoltaic applications are developed to analyse the shading conditions on the ground used for crop production. The models have shown remarkably similar accuracy compared to commercial software such as PVsyst® and SketchUp®. The developed models will help reduce the crop yield uncertainty under agrivoltaic systems by providing accurate photosynthetically active radiation distribution at the crop level. The distribution was further analysed using a light homogeneity index and calculating the yearly photosynthetically active radiation reduction. The homogeneity and photosynthetically active radiation reduction varied significantly depending on the agrivoltaic system design, from 91% to 95% and 11% to 34%, respectively. To identify the most suitable agrivoltaic system layout dependent on crop and geographical location, it is of fundamental importance to study the effect of shadings with distribution analysis.

## 1 Introduction

Shadings on photovoltaic (PV) modules cause significant losses in electricity production, especially in cases where there are no diodes, as the performance is decreased approximately proportionally to the irradiance reduction [1]. Therefore, several research activities have been focused on developing methods to accurately estimate the yield loss of solar arrays surrounded by objects or due to self-shading when arranged in multiple rows. Shading calculations for solar arrays have been developed and used for several years. In 1987, Bany & Appelbaum [2] developed equations to calculate the shadow that occurs on a field of solar collectors during the day. It is common to have more complex situations than a field of solar collectors. Hence, Cascone et al. [3] developed a calculation procedure of the shading factor under complex boundary conditions that can be implemented for every sky condition. The procedure can obtain the instantaneous, daily average, or monthly average shading factor values. Nowadays, several commercial software solutions are used to calculate the shading factors, but these products tend to have limitations. For instance, some commercial products do not provide the flexibility and control by changing the timesteps or angular displacement needed for shading factors calculations. To provide more flexibility, control, and transparency over the shading factor calculation process, Silva et al. [4] created a model in Matlab® to calculate the beam and diffuse shading factors. Their model was validated with PVsyst® and System Advisor Model (SAM)® using obstacles shading the fixed tilted PV array. Their model has shown similar results as SAM®. However, the PVsyst® result differed significantly from their model, with an $R^2$ of 46% when analysing one module shaded by one infinite row. The shading factor simulation over a long period requires high computational power. To reduce the time complexity, Melo et al. [5] developed a method to estimate the shading factor and irradiation received on a surface, using a bi-linear interpolation. The simulation time could be reduced from 955 seconds to 39 seconds without reducing the shading factor accuracy.

Even though shading factors are used frequently for PV systems performance assessment, there is still a limited amount of research regarding new emerging PV systems layouts, especially for agrivoltaic applications that have the possibility of reducing the competition of land used for either food or electricity production.

Agrivoltaic systems can increase renewable energy capacity and produce food simultaneously [6], be a climate adaptation technology, and support society to meet the energy demand by only using a single per cent of global agricultural land [7]. Agrivoltaic systems have been found to also have a great synergy with animals by simultaneously using shading as a shelter and the land for grazing [8]. Additionally, there are several mutual benefits, such as reduced PV module heat stress, reduction of plant drought stress, and increased food production under certain climatic conditions [9]. Despite all the benefits, a tremendous lack of knowledge exists on how the properties of the crop (i.e., yield, energy content, and morphology) underneath the PV modules are affected due to shadings [9–11]. Therefore, accurate shading calculations on the ground are necessary to understand better the effects of combining PV systems and agriculture. Campana et al. [12] analysed a vertically mounted agrivoltaic system located in Sweden. In their study, diffuse and beam shading factors are calculated to get the photosynthetically active radiation (PAR) distribution on the ground. The PAR is needed for the crop model Environmental Policy Integrated Climate (EPIC) developed by Williams et al. [13] to get the actual crop yield. The model has shown to be robust in identifying crop productivity, especially in open-field conditions [12]. Amaducci et al. [14] developed a shading and radiation model to compute the crop yield for an agrivoltaic system using two-axis trackers in North Italy. Their model indicates that agrivoltaic systems support clean energy conversion, crop yield, and water saving. Nevertheless, there is still a lack of studies analysing the shading that occurs under the panels of an agrivoltaic system, and this is a knowledge gap that must be filled.

In this paper, a model is developed to accurately calculate the PAR distribution reaching the crops for fixed, one-axis, and two-axis agrivoltaic systems. In agrivoltaic systems, using either one- or two-axis trackers is common to control the irradiance on the crops grown underneath the PV system [15]. The model can be employed as a starting point for accurately estimating crop yields at any given location under the most common agrivoltaic system designs. The schematic of the modelling chain is presented in Figure *1*.

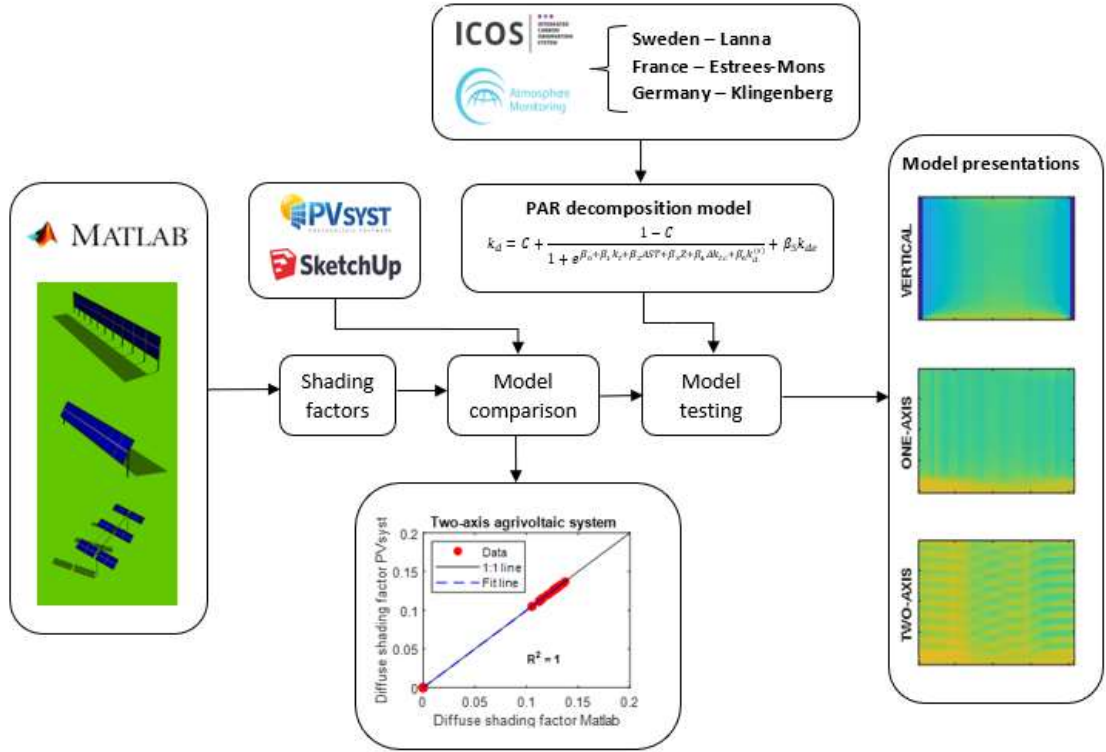

Figure 1 Modelling schematic

The paper is structured as follows. Section 2 describes the methodology used to calculate the beam and diffuse shading factors for the fixed, one-axis, and two-axis agrivoltaic systems. Section 3 presents the model comparisons and PAR distribution at different locations. Section 4 summarizes the results of this study.

## 2 Methodology

The shadow effect on the solar modules includes the direct and diffuse shading factors. The direct component depends on the module geometry, obstacles, and the sun's position. The diffuse component only depends on module geometry and obstacles. There is also a reflected component reduced by obstacles, reducing the albedo [4]. The albedo effect is not included in this work as it is out of scope. The shading factors used to calculate the beam horizontal irradiance (BHI) and the diffuse horizontal irradiance (DHI) can be calculated as follows:

$$BHI_S = BHI(1 - f_b), \tag{1}$$

$$DHI_S = DHI(1 - f_d), \tag{2}$$

where, $f_b$ is the beam shading factor, $f_d$ is the diffuse shading factor, $BHI_S$ is the shaded beam horizontal irradiance, and $DHI_S$ is the shaded diffuse horizontal irradiance. The input parameters used for the vertical, one-axis, and two-axis agrivoltaic system designs are presented in Table 1. Due to construction limitations, the one-axis and two-axis maximum tilt angles are assumed to be -60 and 60 degrees.

Table 1 Design parameters for the vertical, one-axis, and two-axis agrivoltaic systems.

| Input parameters | Vertical | One-axis | Two-axis |
|---|---|---|---|
| Panel width [m] | 1 | 1 | 1 |
| Panel length [m] | 2 | 2 | 2 |
| Number of panels | 40 | 40 | 40 |
| Total panel area [m²] | 80 | 80 | 80 |
| Number of rows | 2 | 2 | 2 |
| Row spacing [m] | 10 | 10 | 10 |
| Row length [m] | 20 | 20 | 20 |
| Crop area [m²] | 200 | 200 | 200 |
| Pitch [m] | - | - | 2 |
| Height [m] | 0 | 3 | 3 |
| Fixed tilt angle [°] | 90 | - | - |
| Azimuth angle [°] | 0 | 0 | 0 |
| Maximum tilt angle [°] | - | 60 | 60 |
| Minimum tilt angle [°] | - | -60 | -60 |

The coordinates of the PV module are calculated with simple trigonometric relationships as follows:

$$P1 = \begin{bmatrix} x_0 \\ y_0 \\ z_0 \end{bmatrix}, \tag{3}$$

$$P2 = \begin{bmatrix} x_0 \\ y_0 + W \\ z_0 \end{bmatrix}, \tag{4}$$

$$P3 = \begin{bmatrix} x_0 + L \\ y_0 \\ z_0 \end{bmatrix}, \tag{5}$$

$$P4 = \begin{bmatrix} x_0 + L \\ y_0 + W \\ z_0 \end{bmatrix}, \tag{6}$$

where, $x_0$, $y_0$, and $z_0$ are the initial PV module coordinates located in the plane, L is the length of the panel (m), and W is the width of the panel (m). The tilt angle $\omega_{ITC}$ (°) for both fixed and one axis tracking PV systems follows the solar elevation with a rotation around the y-axis. The tilt angles are calculated with the equations defined by Lorenzo et al. [16]. First, an ideal tracking angle $\omega_{IT}$ including mutual shadowing must be calculated and is given by:

$$\tan(\omega_{IT}) = \frac{x}{z}, \tag{7}$$

where x, y, and z are the Cartesian coordinates of the Sun, which is referred to as a reference system with the z-axis pointing to the zenith, the y-axis pointing south, and the x-axis pointing west. The components of the solar vector are given as follows [16]:

$$\mathbf{s} = \begin{cases} x = \cos(\alpha_s)\sin(\gamma_s) \\ y = \cos(\alpha_s)\cos(\gamma_s), \\ z = \sin(\alpha_s) \end{cases} \tag{8}$$

where, $\gamma_s$ is the solar azimuth angle (°), and $\alpha_s$ is the solar elevation angle (°). The solar vector components are calculated according to a panel-oriented coordinate system, which is defined with a north-south axis direction. To adjust the system deviations for different azimuth angles $\alpha_Z$ relative to the N-S direction and axis tilts from the horizontal plane angle $\beta_Z$, the system has to be referred to as follows:

$$\begin{cases} x' = x\cos(\alpha_Z) - y\sin(\alpha_Z) \\ y' = x\cos(\beta_Z)\sin(\alpha_Z) + y\cos(\beta_Z)\cos(\alpha_Z) - z\sin(\beta_Z), \\ z' = x\sin(\beta_Z)\sin(\alpha_Z) + y\sin(\beta_Z)\cos(\alpha_Z) + z\cos(\beta_Z) \end{cases} \tag{3}$$

The shaded fraction SF and shadow length s are given by:

$$SF = \max\left[0, \left(1 - \frac{L_{EW}}{s}\right)\right], \tag{9}$$

$$s = \frac{1}{\cos(\omega_{IT})}, \tag{10}$$

where, $L_{EW}$ is the distance between axes from east to west. In Figure 2 the geometry of a one-axis horizontal tracking is depicted.

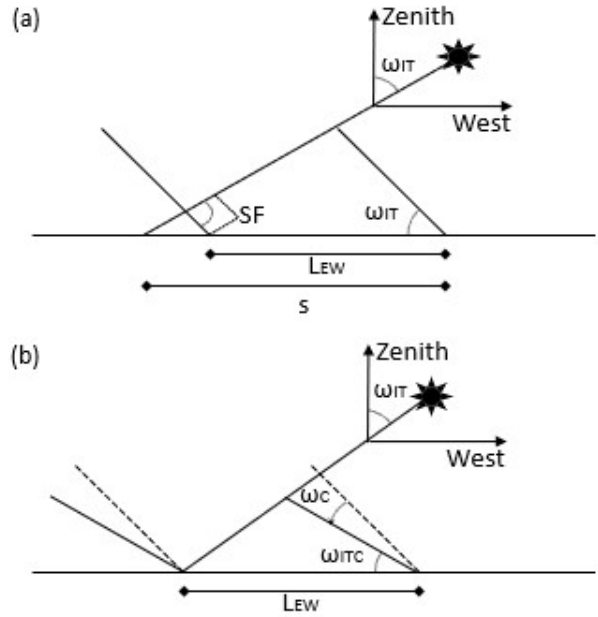

Figure 2 One-axis horizontal N-S oriented tracking. (a) ideal tracking with mutual shadowing, (b) corrected ideal tracking (back-tracking). Depicted from (Lorenzo et al. [16]).

The ideal tracking angle can be used to calculate the corrected ideal tracking angle $\omega_{ITC}$ and is given by:

$$\omega_C = \omega_{IT} - \omega_{ITC}, \tag{11}$$

$$\cos(\omega_C) = L_{EW} \cos(\omega_{IT}), \tag{12}$$

where, $\omega_C$ is the back-tracking correction angle. For a fixed PV system, the tilt angle is fixed. The rotation around the y-axis is done by using the rotation matrix defined as follows [17]:

$$R_y = \begin{bmatrix} \cos(\omega_c) & 0 & \sin(\omega_c) \\ 0 & 1 & 0 \\ -\sin(\omega_c) & 0 & \cos(\omega_c) \end{bmatrix}, \tag{13}$$

The two-axis tracking PV system is rotated in the y-direction similarly as done for one-axis systems. In Figure *3*, the geometry of a two-axis horizontal tracker is presented.

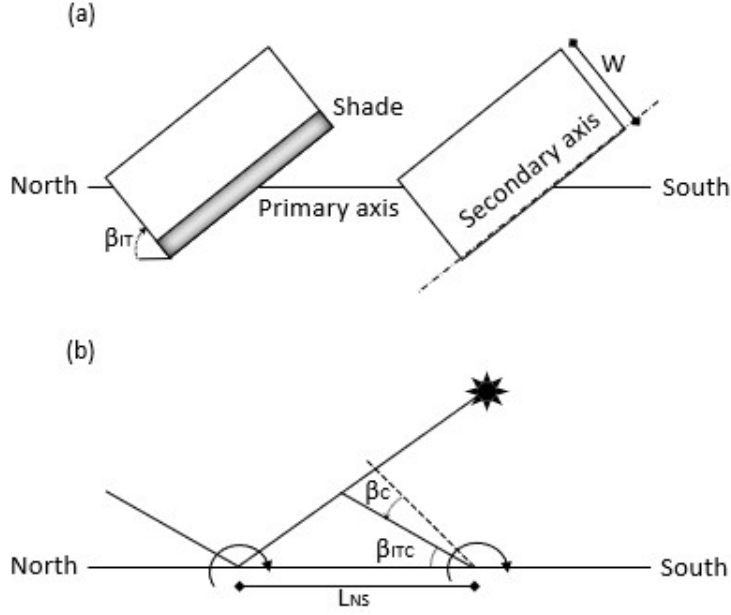

Figure 3 Two-axis horizontal N-S oriented axis tracking, (a) ideal tracking with mutual shading, (b) corrected ideal tracking (back-tracking). Depicted by Lorenzo et al. [16].

However, the second rotation follows the solar azimuth angle and the ideal tracking angle $\beta_{IT}$ for the x-axis is given by:

$$\tan(\beta_{IT}) = \frac{y}{(x^2 + y^2)^{1/2}}, \tag{14}$$

The corrected ideal tracking angle $\beta_{ITC}$ is given by:

$$\cos(\beta_{ITC}) = \frac{L_{NS}}{W} \cos(\beta_{IT}), \tag{15}$$

where, $L_{NS}$ is the distance between the axes from north to south, and W is the panel width. The second rotation is done by using the rotation matrix around the x-axis as follows [17]:

$$R_x = \begin{bmatrix} 1 & 0 & 0 \\ 0 & \cos(\omega_{1c}) & -\sin(\omega_{1c}) \\ 0 & \sin(\omega_{1c}) & \cos(\omega_{1c}) \end{bmatrix}, \tag{16}$$

The final position of the coordinates of the PV modules is obtained after the rotation around both the y-axis and x-axis and one translational displacement. The rotations are performed by multiplying each panel point by a rotation matrix. The translational displacement is performed by first subtracting the current position of the panel's coordinate with rotation center points. The mathematical expression of the rotations and translational displacement is given by [18]:

$$P_i = R_x R_y \begin{bmatrix} P_{ix} - x_c \\ P_{iy} - y_c \\ P_{iz} - z_c \end{bmatrix} + \begin{bmatrix} x_c \\ y_c \\ z_c \end{bmatrix}, \quad i \in \{1,2,3,4\}, \tag{17}$$

where, $P_i$ are the points defining the PV module, $P_{ix}$, $P_{iy}$, and $P_{iz}$ are the coordinates in the x-axis, y-axis, and z-axis respectively for the PV module points; $x_c$, $y_c$, and $z_c$ are the coordinates describing the centre of the PV module. The equation of the normal vector $n$ can be retrieved from the orientation and the tilt of the PV modules plane [3]:

$$\boldsymbol{n} = \begin{cases} a = \cos(\gamma) \sin(\delta) \\ b = \sin(\gamma) \sin(\delta), \\ c = \cos(\gamma) \end{cases} \tag{18}$$

Once the normal vector $n$ and solar vector $s$ are defined, the projection of the PV module points that define the shading on the desired plane can be solved by calculating the straight-line projections and can be determined as follows [3]:

$$P' = \begin{cases} P_x + x \cdot t \\ P_y + y \cdot t, \\ P_z + z \cdot t \end{cases} \tag{19}$$

The coordinates of point P', projection onto the plane along the straight line are given by the intersection of the straight line and the plane and can be solved by calculating parameter $t$ [3]:

$$t = -\frac{aP_x + bP_y + cP_z}{ax + by + cz}, \tag{20}$$

The beam shading factor can be calculated as follows:

$$f_b = \frac{A_{shade}}{A_{tot}}, \tag{21}$$

where $A_{shade}$ is the shaded area (m²) and $A_{tot}$ (m²) is the total reference area for the crop between two PV module rows. In this study, an assumed crop area of 200 m², the area between the PV rows is used to define the total reference area. The total shading on the reference area for multiple panels is calculated by calculating the union and is given by:

$$A_{shade} = \bigcup_{i=1}^{n} A_{shade,i}, \tag{22}$$

where, n is the number of PV modules, and $A_{shade,i}$ is the shaded area from each respective PV module. A shading table for the direct component is used to discretize the sky dome with altitude angles from 0° to 90° and azimuth angles from −180° to +180° of 1 degree. The diffuse shading factor $f_d$ is calculated using the approach used in [3,12,19] and is given by:

$$f_d = \frac{\int_{\alpha=0}^{\frac{\pi}{2}} \int_{\gamma=0}^{2\pi} f_b R_{\alpha\gamma} \cos\theta d\Omega}{\int_{\alpha=0}^{\frac{\pi}{2}} \int_{\gamma=0}^{2\pi} R_{\alpha\gamma} \cos\theta d\Omega}, \qquad (23)$$

where $\alpha$ and $\gamma$ are the sun altitude and azimuth angles, $R_{\alpha\gamma}$ is the radiance (W/m²sr), $\theta$ is the angle of incidence (°), which is the angle between the sky element and the normal to the surface, and $\Omega$ is the solid angle (sr). The solid angle can also be expressed as $\partial\Omega = \cos(\alpha) \cdot \partial\alpha\partial\gamma$. To simplify the diffuse shading factor calculation and remove radiance $R_{\alpha\gamma}$, an isentropic sky has to be considered. The discretised sky dome can be numerically computed by:

$$f_d = \frac{\sum_{i=1}^{\alpha} \sum_{j=1}^{\gamma} f_{b_{ij}} \cos(\theta_{ij}) \cos(\alpha_i)}{\sum_{i=1}^{\alpha} \sum_{j=1}^{\gamma} \cos(\theta_{ij}) \cos(\alpha_i)}, \qquad (24)$$

### 2.1 Fixed vertical PV

The fixed vertical agrivoltaic system PV panel geometry is defined by four points $P_i$ in the plane that corresponds to the vertexes of the PV module or array. Once those points are defined, the projections of those vertices on the horizontal plane can be used to calculate the shaded area that occurs on the ground at a given time step. It is assumed that grass growing close to the supporting structure does not allow the sunlight to pass through similar to Campana et al. [12]. An illustration of the fixed vertical agrivoltaic system is depicted in Figure 4.

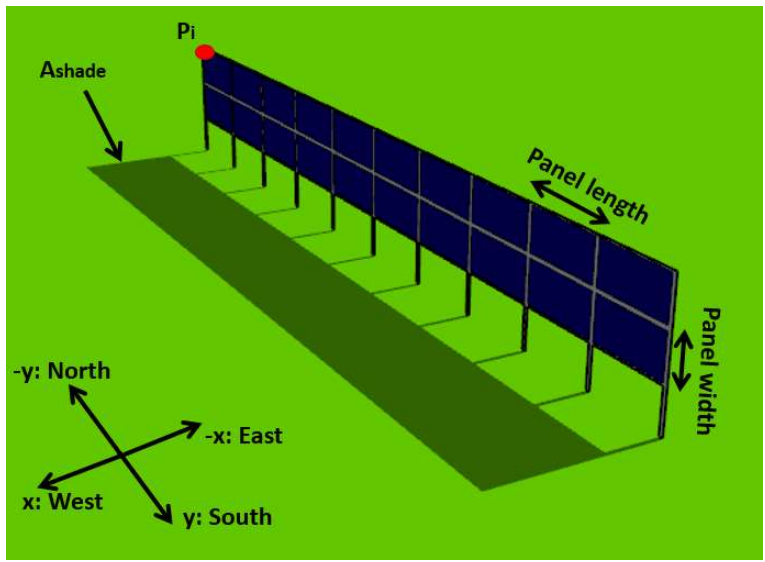

Figure 4 Vertical fixed agrivoltaic system.

## 2.2 One-axis and two-axis tracking

In this study, one-axis and two-axis tracking and back-tracking angles have been calculated using similar equations as Lorenzo et al. [16]. The coordinates for the PV panel can be defined as done previously for the fixed vertical PV. An illustration of the one-axis and two-axis tracking systems used in this study is presented in Figure 5, and Figure 6, respectively.

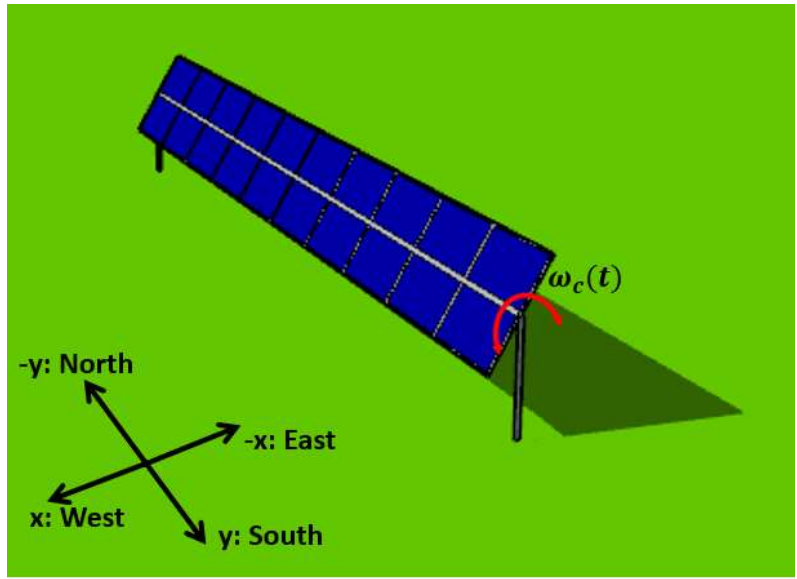

Figure 5 One-axis tracking agrivoltaic system.

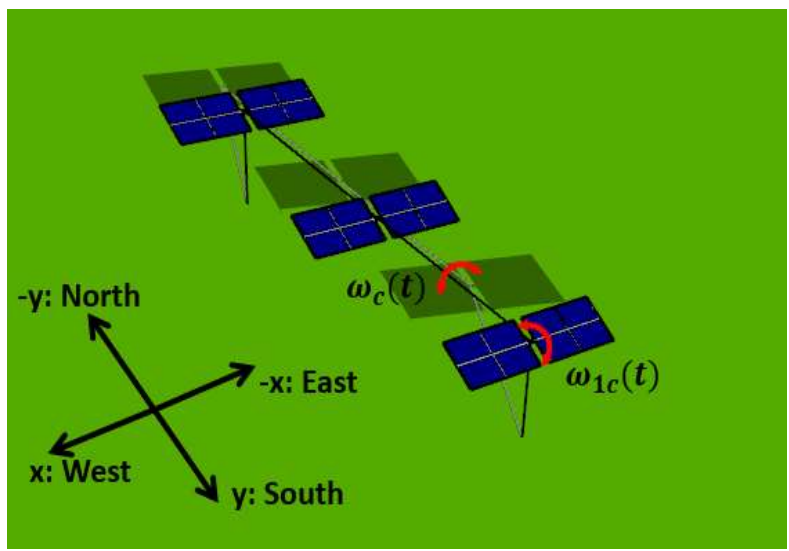

Figure 6 Two-axis tracking agrivoltaic system.

## 2.3 Model comparison

Two commercial software are used to compare the shading factors of the proposed model: PVSyst® and SketchUp®. To make the comparison fair, the design of the three studied systems modelled in PVSyst® and SketchUp® are identical to the dimensions described in Table 1. The beam and diffuse shading factors have been calculated for four representative days, one for each season (i.e., the equinox of spring, the solstice of summer, the equinox of autumn, and the solstice of winter). The shading factor comparison refers to Kärrbo Prästgård (Latitude 59.6099°N, Longitude 16.5448°E, Altitude 20m), Västerås, Sweden. In this location, the first agrivoltaic system in Sweden was installed in 2021. The system is currently installed on a pasture grass field. The data is gathered from Meteonorm software, providing monthly meteorological data for any location on the earth. Using stochastic models, it generates synthetic hourly values from the monthly values [20]. The time resolution for validation is 1-hour. The methodology employed to obtain the shading factors from both products is briefly presented in this section.

### 2.3.1 SketchUp®

SketchUp® is a 3D geometry modelling tool used by architects, engineers, and designers. The tool contains a real-time shadow engine that can be used, for instance, for sun exposure analysis. In this study, SketchUp® Make 2017 is used with the TIG Shadow Projector v7.0 [21] plugin to obtain the amount of shading on a surface for a specific time and location. Consequently, the beam shading factors for the three agrivoltaic configurations are derived from this tool. The real-time shadow engine from SketchUp® simulates the sun's position at a specific time of the year. Therefore, the engine visualizes the shadings on the ground created by the solar modules of the different agrivoltaic configurations. Except for the vertical configuration, which is fixed, the geometries (rotation angles) of the one-axis and two-axis agrivoltaic systems are manually adjusted at every time step according to the optimized values from the tracking systems. Figure 7 shows the schematics of the different systems in

SketchUp® and an example of the results displayed by the plugin. Concerning the diffuse shading factor, unfortunately, the validation could not be performed using SketchUp®.

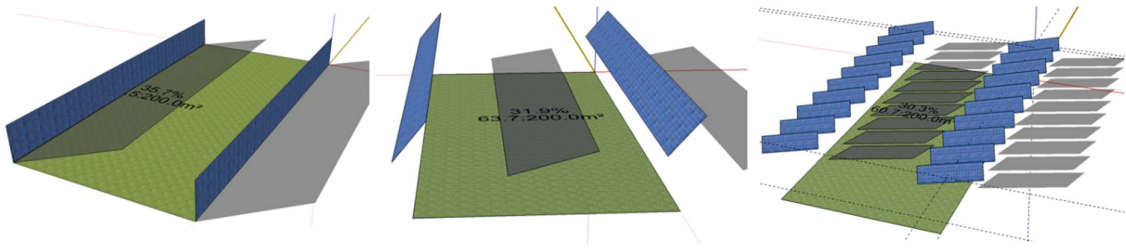

Figure 7 Schematics using SketchUp® Make 2017 and the plugin TIG Shadow Projector of the three AV systems: vertical (left), one-axis tracking (middle), and two-axis tracking (right). The yellow lines point towards the true North. The area of analysis is the grass-coloured rectangle.

### 2.3.2 PVsyst®

PVsyst® is one of the most powerful software to model and simulate PV systems and is designed for architects, researchers, and engineers [22]. The software has a 3D shading scene to analyse the shadings from a specific PV system design and shading scene. In this study, the 3D shading scene was used to calculate both the diffuse shading factor and beam shading factor for the three agrivoltaic configurations. In PVsyst®, to calculate the shading factors on the ground instead of on the PV array as it is conventionally performed, the scene had to be designed in a specific way as the software does not have any agrivoltaic/ground shading scene solution. The crop area was covered with a horizontal PV module corresponding to the grass, and the vertical and tracker modules were converted to shading objects. The shading factors can be calculated hourly by re-constructing the shading scene for the trackers to adjust their tracking position. Figure 8 shows a schematic of the different shading scenes in PVsyst®.

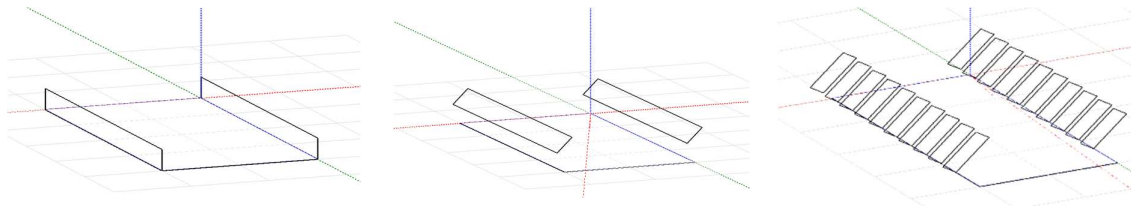

Figure 8 Schematics using PVsyst® Shading scene of the three AV systems: vertical (left), one-axis tracking (middle), and two-axis tracking (right). The red line point towards the true East.

## 2.4 Model testing

The model developed in Matlab® is tested at three different locations in Europe with all three developed designs. The test is conducted by calculating the yearly PAR distribution at these sites. The data on PAR and GHI are collected from the Integrated Carbon Observation System (ICOS) [23]. The locations were chosen based on the availability of data at cropland sites. The locations with available hourly data used in this paper are presented in Table 2.

Table 2 Data of ICOS stations.

| Station | Latitude (°) | Longitude (°) | Elevation (m) | Year |
|---|---|---|---|---|
| Lanna, Sweden | 58.33 | 13.1 | 75 | 2018 |
| Estrees-Mons, France | 49.87 | 3.02 | 85 | 2018 |
| Klingenberg, Germany | 50.89 | 13.52 | 478 | 2018 |

The PAR is decomposed into its diffuse and direct components. This decomposition of PAR is critical in agrivoltaic systems due to the variable shadings caused by the panels on the crops, creating a non-homogenous PAR/diffuse PAR distribution during the day [24]. The decomposition model used in this study is YANG2 and is given by [25]:

$$k_d = C + \frac{1-C}{1+e^{\beta_0+\beta_1 k_t+\beta_2 AST+\beta_3 Z+\beta_4 \Delta k_{tc}+\beta_6 k_d^{(s)}}} + \beta_5 k_{de}, \tag{25}$$

$$k_t = \frac{GHI}{E_{ext}}, \tag{26}$$

$$\Delta k_{tc} = \frac{G_{cs}}{E_{ext}} - k_t, \tag{27}$$

$$k_{de} = \max\left(0, 1 - \frac{G_{cs}}{GHI}\right), \tag{28}$$

where the model coefficients used in this study have been fitted to three ICOS stations from Sweden [24] being C = 0.0888, $\beta_0 = -26258$, $\beta_1 = 7.2506$, $\beta_2 = -0.0458$, $\beta_3 = 0.0099$, $\beta_4 = -0.0839$, $\beta_5 = 0.5002$, $\beta_6 = -2.1731$. $k_d^{(s)}$ is the satellite-derived diffuse fraction, $G_{cs}$ is the clear sky GHI [W/m²], AST is the apparent solar time [h], $E_{ext}$ is the extraterrestrial radiation [W/m²], and $k_t$ is the clearness index. The $k_d$ value from the YANG2 model can be used in the relationship defined by Spitters et al. [26] to get the correct PAR estimates:

$$k_d^{PAR} = \frac{PAR_{diffuse}}{PAR_{total}} = \frac{[1+0.3(1-(k_d^{YANG2})^2)]k_d^{YANG}}{1+(1-(k_d^{YANG2})^2)\cos^2(90-\beta)\cos^3(\beta)}, \tag{29}$$

where, $\beta$ is the solar elevation angle [°]. The satellite data is obtained through the Copernicus Atmosphere Monitoring Service radiation service (CAMS) which has a spatial coverage of −66° to 66° in both latitudes and longitudes and a temporal resolution of 15-min [27]. The ground area is discretized by 0.25m × 0.25m. The discretization makes it possible to analyse the variation of PAR using different agrivoltaic system designs at various locations. To compare different agrivoltaic systems, a light homogeneity index (LHI) can be used, and is defined as the ratio of sample standard deviation to mean and is given by the following equation:

$$LHI = 100 \cdot \left(1 - \frac{\frac{1}{n-1}\sum_{n-1}^{n}(x_i - \bar{x})^2}{\bar{x}}\right), \tag{30}$$

where, $n$ is the number of discretized areas on the ground, $x_i$ is the total PAR including shadings from the agrivoltaic system, $\bar{x}$ is the average yearly PAR. An LHI of 100% corresponds to a PAR light distribution that is equal throughout the ground area and can in other terms be defined as a complete homogenous light distribution.

## 3   Results and discussion

In this section, the shading factor validation is first presented and is then followed by presenting the PAR distribution for the vertical, one-axis, and two-axis agrivoltaic systems in Lanna, Estrees-Mons, and Klingenberg. Figure 9 presents the validation results for the vertical, one-axis, and two-axis agrivoltaic systems for all days.

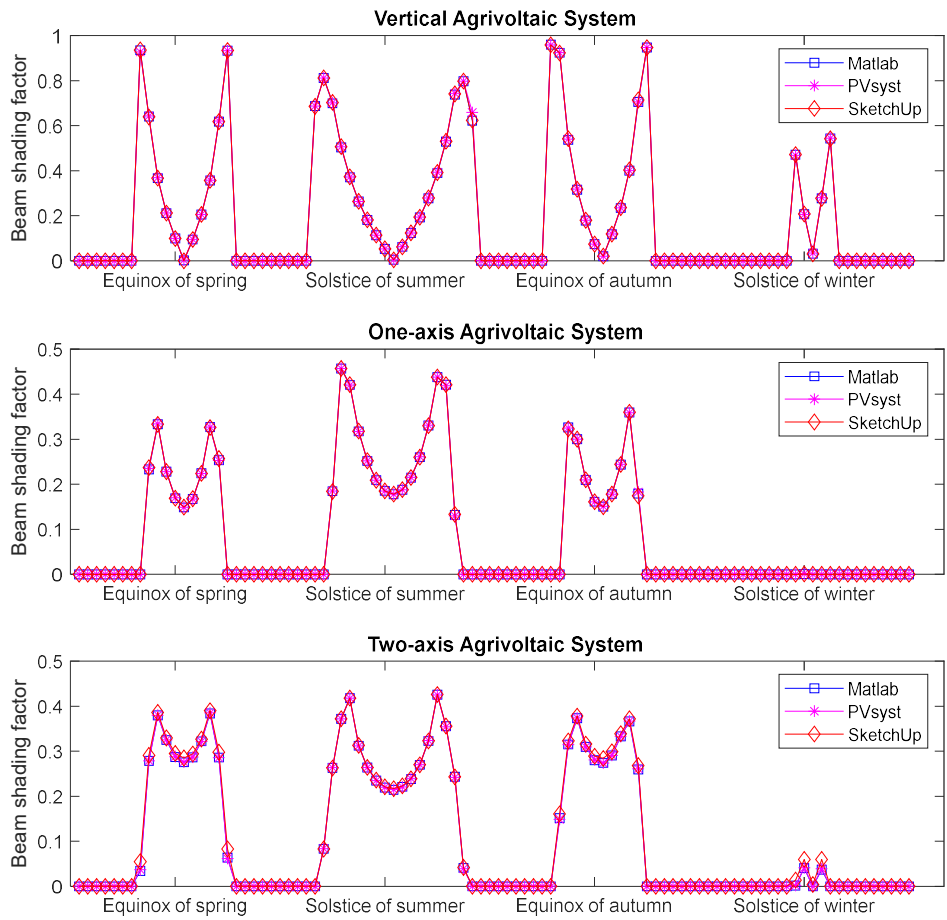

Figure 9 Beam shading factor time series for vertical, one axis, and two axis agrivoltaic systems for four representative days. Comparison between the Matlab® model developed in this study and the models developed in PVsyst® and SketchUp®.

The MBE and RMSE for the beam shading factors are summarized in Table 3. The Matlab® model shows a good correlation with both PVsyst® and SketchUp® in all the investigated agrivoltaic layouts models.

Table 3 Beam shading factor error metrics between Matlab®, PVsyst®, and SketchUp®.

| System | Beam shading factor | | | |
|---|---|---|---|---|
| | PVsyst® | | SketchUp | |
| | MBE | RMSE | MBE | RMSE |
| Vertical | 0.0008 | 0.0040 | 0.0011 | 0.0039 |
| One-axis | 0 | 0.0006 | 0 | 0.0012 |
| Two-axis | 0.0003 | 0.0006 | 0.0026 | 0.0055 |

In Figure 10, the scatter plots illustrate the comparison between the Matlab®, PVsyst®, and SketchUp® models. The $R^2$ values for all models have shown a performance above 99%, which shows that the Matlab® models have a good correlation with commercially available products.

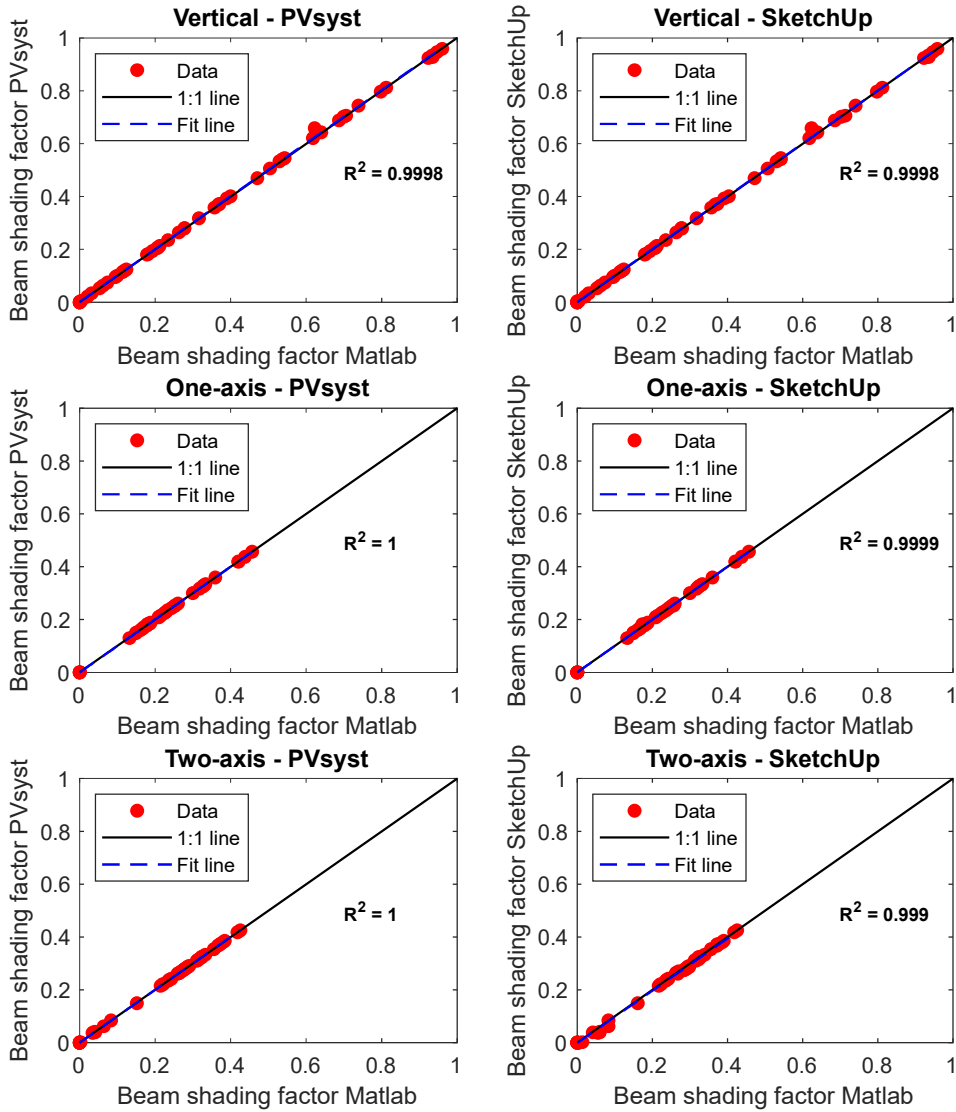

Figure 10 Scatter plots for beam shading factor. Comparison between the Matlab® model developed in this study and the models developed in PVsyst® and SketchUp®.

For the assumptions made concerning the isotropy of the sky, the diffuse shading factor must be determined only once for the fixed tilt agrivoltaic system layout. Nevertheless, for the one-axis and two-axis agrivoltaic systems layouts, the diffuse shading factor must be determined hourly due to the involved rotations. Figure 11 presents the diffuse shading factor time series for the one- and two-axis agrivoltaic systems for all the representative days. The diffuse shading factors from the Matlab® model are accurately calculated compared to PVsyst®. The diffuse shading factor was obtained by manually adjusting the near shading scene in PVsyst®.

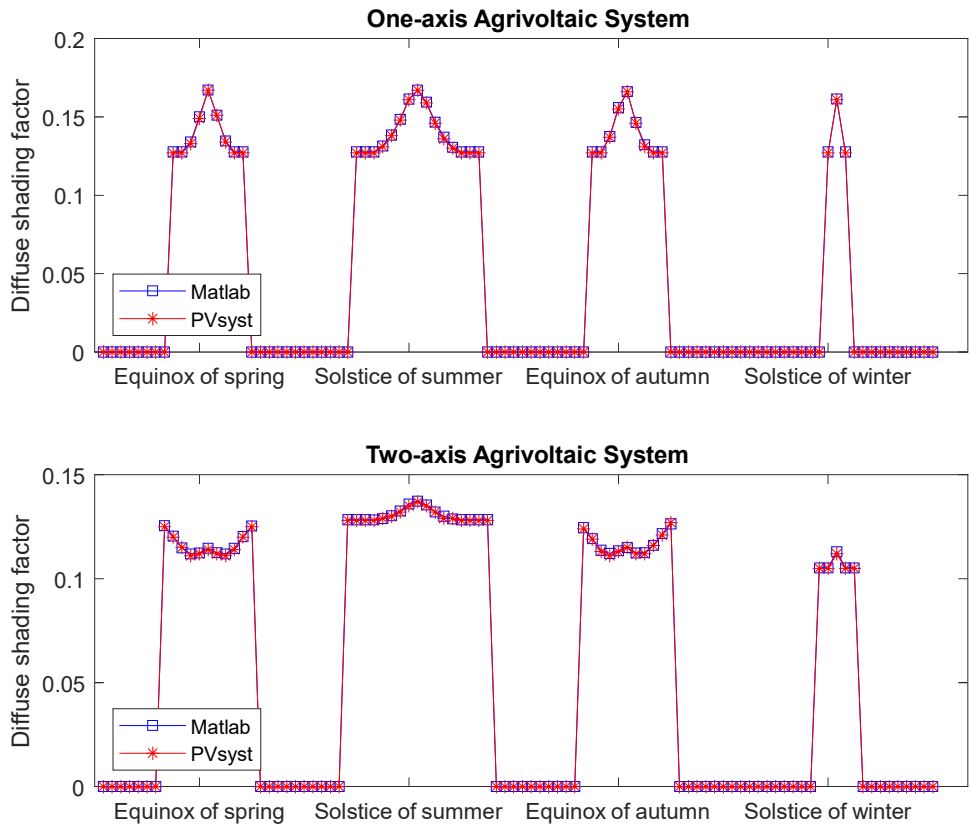

Figure 11 Diffuse shading factor time series for one axis and two axis agrivoltaic systems. Comparison between the Matlab® model developed in this study and the model developed in PVsyst®.

The MBE and RMSE for the diffuse shading factors are summarized in Table 4. The model developed in Matlab® is estimating the diffuse shading factors accurately for all four representative days, for both the one-axis and two-axis tracker as seen in Figure 11.

Table 4 Diffuse shading factor error metrics between the model developed in this study in Matlab® and PVsyst®.

| System | Diffuse shading factors PVsyst® | |
|---|---|---|
| | MBE | RMSE |
| Vertical | 0.0018 | 0.0018 |
| One-axis | −0.0002 | 0.0004 |
| Two-axis | −0.0002 | 0.0004 |

In Figure 12, the scatter plots illustrate the comparison between the Matlab® and PVsyst® for one-axis and two-axis agrivoltaic systems. The $R^2$ shows a good correlation between all models in Matlab® and PVsyst®.

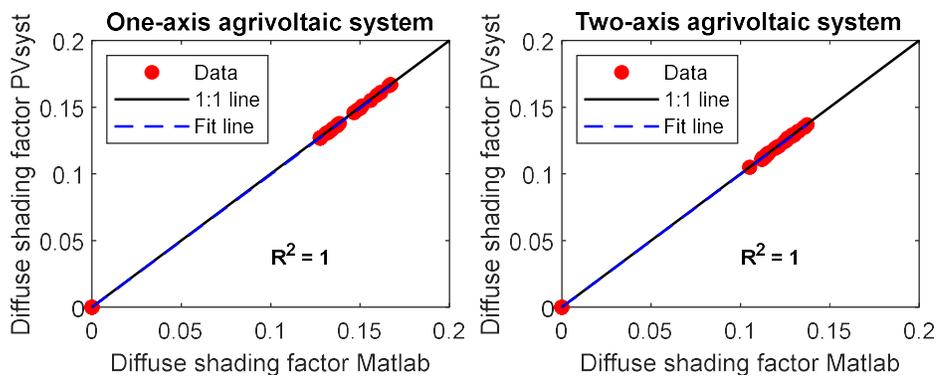

Figure 12 Scatter plots for diffuse shading factor. Comparison between the Matlab® model developed in this study and the model developed in PVsyst® one-axis and two-axis agrivoltaic systems.

The validation of the developed model in Matlab® has shown a good correlation and can therefore be used for accurate shading calculations under agrivoltaic systems. Figure 13 presents the yearly PAR distribution for the three different agrivoltaic configurations developed in this study for Lanna in Sweden, Estrees-Mons in France, and Klingenberg in Germany.

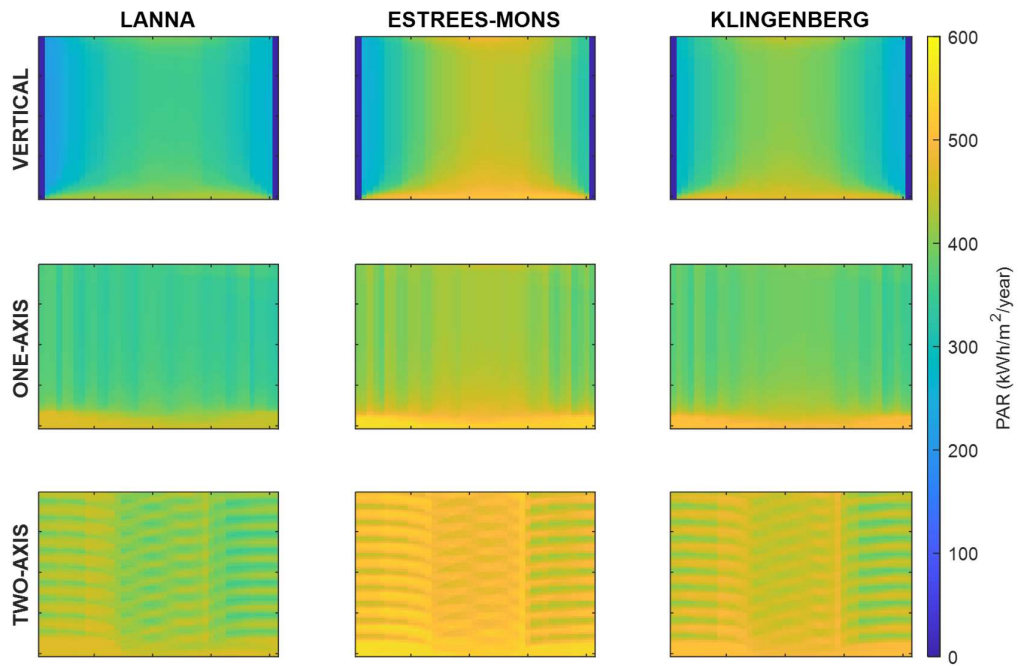

Figure 13 Yearly PAR distribution for the vertical, one-axis, and two-axis designs for Lanna in Sweden, Estrees-Mons in France, and Klingenberg in Germany.

The PAR distribution can be used to analyse several agrivoltaic systems layouts to identify the most suitable layout for a specific crop or crop rotation at any location. The model can be deployed to reduce crop yield simulation uncertainties under agrivoltaic systems significantly. As seen in Figure 11, the PAR varies depending on the choice of system design and geographical location. The two-axis tracking system showed the highest yearly PAR. For instance, in Estrees-Mons the mean PAR was 495 [kWh/m2, year]. The yearly mean PAR varies significantly dependent on system design. In Estrees-Mons, the difference in yearly mean PAR between the two-axis and the vertical system was 117 [kWh/m2, year]. It is also essential to have a low variation of PAR throughout the reference crop area, as a high variation of PAR could heavily affect the crop yield.

Additionally, crop water stress can be increased due to PAR heterogeneity throughout the field. To avoid a system design with high PAR heterogeneity, the LHI is used for all three system designs to assess PAR homogeneity. In Table 5, the LHI for the vertical, one-axis, and two-axis designs are presented. 100% corresponds to a homogenous distribution and 0% corresponds to a non-homogenous distribution.

Table 5 Light homogeneity index for the vertical, one-axis, and two-axis designs for Lanna in Sweden, Estrees-Mons in France, and Klingenberg in Germany.

|  | *LHI* | | |
| --- | --- | --- | --- |
|  | **Lanna** | **Estrees-Mons** | **Klingenberg** |
| **Vertical** | 92.68% | 93.76% | 93.48% |
| **One-axis** | 91.83% | 92.03% | 91.84% |
| **Two-axis** | **95.25%** | **95.39%** | **95.26%** |

All three designs had a high LHI above 91% at all locations. However, the system design with the highest LHI was the two-axis system with an LHI above 95% at all three locations and at the same time having the highest yearly PAR reaching the crop area. The LHI gives a better understanding of the variation of PAR over the field. However, it is still essential to know the total reduction of PAR for the specific system as a complement to the LHI (Table 6). The total yearly PAR reduction can be calculated by fractioning the PAR without shading on the ground and the PAR with the respective agrivoltaic system.

Table 6 Total yearly PAR reduction for the vertical, one-axis, and two-axis designs for Lanna in Sweden, Estrees-Mons in France, and Klingenberg in Germany.

|  | *Total yearly PAR reduction* | | |
| --- | --- | --- | --- |
|  | **Lanna** | **Estrees-Mons** | **Klingenberg** |
| **Vertical** | 34.72% | 32.12% | 32.24% |
| **One-axis** | 22.46% | 23.14% | 23.01% |
| **Two-axis** | **11.41%** | **11.17%** | **11.82%** |

The yearly PAR reduction varies significantly depending on the choice of agrivoltaic design. The choice of an agrivoltaic system is fundamental as the shading significantly impacts the amount of yearly PAR reaching the crops. In this case, the variation of yearly PAR reduction could vary from 11% to 34%, dependent on the choice of agrivoltaic system. It can be noticed that the two-axis agrivoltaic system had the lowest yearly PAR reduction at all studied sites. However, a low yearly PAR reduction does not necessarily represent an optimal agrivoltaic system design, as the crop's photosynthetic rate is a function of light intensity which varies between species. This shows that it is important to accurately design the agrivoltaic system by finding the optimal synergy between the crop and PV systems. The Matlab® model developed in this study can be used as a universal model for agrivoltaic systems at any location in the world and gives the freedom to analyze shading factors and how these affect the PAR distribution at the crop level and consequently the crop yield for any agrivoltaic system. Additionally, the model in the study can be integrated into modelling platforms such as those developed by Campana et al. [12] and Amaducci et al. [14] to maximize crop yield and electricity production, and PV modules orientation, especially for those agrivoltaic systems layouts equipped with tracking systems.

## 4  Conclusions

In this study, three mathematical models are developed in Matlab® for accurately calculating the beam and diffuse shading factors, which can be used to estimate the shaded beam horizontal irradiance and shaded diffuse horizontal irradiance at a ground level. This modelling effort has been carried out to accurately depict the PAR distribution and reduction at ground level, which is of fundamental importance in assessing agrivoltaic system effects on crop yield. The three investigated models were for fixed vertical, one-axis tracking, and two-axis tracking PV systems for agrivoltaic applications. The models were validated with the commercial software PVsyst® and SketchUp®. The main conclusions that can be drawn from this article are as follows:

- All mathematical models developed in this study showed high accuracies compared to PVsyst® and SketchUp® and can be considered robust for shading calculations under agrivoltaic systems. The beam shading factor and diffuse shading factor calculated with the developed Matlab® model showed an $R^2$ of 0.99-1 compared to PVsyst® and SketchUp®.
- Besides comparing the hourly model results with commercially available products, in this study, the model development has also focused on the high temporal and spatial resolution of PAR distribution at the crop level. This assessment is of fundamental importance for studying the effects of shading on crops and crop rotation to identify the most suitable agrivoltaic system layout depending on crops and geographical locations. By calculating a light homogeneity index and yearly PAR reduction it could be seen that the light homogeneity and PAR reduction could vary significantly dependent on agrivoltaic system designs from 91% to 95% and from 11% to 34%, respectively. The two-axis agrivoltaic system had the highest light homogeneity and lowest yearly PAR reduction in Estrees-Mons with a light homogeneity index of 95.39% and total yearly PAR reduction of 11.17%.

  The models developed in Matlab® are universal models that can be easily adapted to other type of agrivoltaic system layouts and integrated with agrivoltaic system modelling platforms for the dual assessment of electricity and crop production.

# 5 Acknowledgments

The authors acknowledge the financial support received from the Swedish Energy Agency through the project "SOLVE solar energy research center", grant number 52693-1. The main author also acknowledges the financial support received from European Energy Sverige AB. The authors also acknowledge the financial support received from the Swedish Energy Agency through the project "Evaluation of the first agrivoltaic system in Sweden", grant number 51000-1. The author Pietro Elia Campana acknowledges Formas - a Swedish Research Council for Sustainable Development, for the funding received through the early career project "Avoiding conflicts between the sustainable development goals through agro-photovoltaic systems", grant number FR-2021/0005.